\documentclass[a4paper]{article}
\usepackage{cite}
\usepackage{colortbl}
\usepackage{array}
\usepackage{colortbl}
\usepackage{graphicx}
\usepackage[marginal]{footmisc}
 \usepackage{enumitem}
\usepackage{subfigure}
\usepackage{INTERSPEECH2021}

\title{Oriental Language Recognition (OLR) 2020:  Summary and Analysis}
\name{Jing Li$^1$, Binling Wang$^1$, Yiming Zhi$^1$, Zheng Li$^2$, Lin Li$^2$, Qingyang Hong$^1$, Dong Wang$^3$}

\address{
  $^1$School of Informatics, Xiamen University, China\\
  $^2$School of Electronic Science and Engineering, Xiamen University, China\\
  $^3$Center for Speech and Language Technologies, Tsinghua University, China}
\email{\{lilin, qyhong\}@xmu.edu.cn, wangdong99@mails.tsinghua.edu.cn}

\begin{document}

\maketitle
\begin{abstract}
The fifth Oriental Language Recognition (OLR) Challenge focuses on language recognition in a variety of complex environments to promote its development. The OLR 2020 Challenge includes three tasks: (1) cross-channel language identification, (2) dialect identification, and (3) noisy language identification. We choose $C_{avg}$ as the principle evaluation metric, and the Equal Error Rate (EER) as the secondary metric. There were 58 teams participating in this challenge and one third of the teams submitted valid results. Compared with the best baseline, the $C_{avg}$ values of Top 1 system for the three tasks were relatively reduced by 82\%, 62\% and 48\%, respectively. This paper describes the three tasks, the database profile, and the final results. We also outline the novel approaches that improve the performance of language recognition systems most significantly, such as the utilization of auxiliary information.
\end{abstract}

\section{Introduction}
The Oriental Language Recognition (OLR) Challenge\cite{wang2016ap16,tang2017ap17,2018AP18,tang2019ap19} is an annual competition, which aims to improve the research on multilingual phenomena and develop language recognition technologies.  OLR 2020 challenge\cite{li2020ap20} is the fifth edition of  the OLR challenge, with some extensions in the competition methodology based on the experience of the past four challenges. ﻿The OLR 2020 challenge includes more languages, dialects and real-life data, and ﻿focuses on more practical and challenging tasks: (1) cross-channel language identification (LID), inherited from OLR 2019 challenge, (2) dialect identification, introduced to OLR challenge for the first time, and (3) noisy LID, also newly introduced this year, considering the general importance of noisy speech processing.

Cross-channel LID is a close-set identification task, which means the language of each utterance is among the known 6 target languages (Cantonese, Indonesian, Japanese, Russian, Korean and Vietnamese), but utterances were recorded with different channels. In real-world applications, channel differences always severely deteriorate the language identification performance. Channel compensation algorithms, such as Linear Discriminant Analysis (LDA)\cite{izenman2013linear} and Probabilistic Linear Discriminant Analysis (PLDA)\cite{ioffe2006probabilistic}, are the most commonly used approaches to address the channel variance, but might be not sufficient. This challenge aims to explore more possibilities in solving the cross-channel issue.

Open-set dialect identification is an open-set identification task, in which three non-target languages (Mandarin, Malay and Thai) and three target dialects (Hokkien, Sichuanese, Shanghainese) in China are merged to form the test set. This combined dialect identification,  with the open-set LID issue, which encourages exploring algorithms that improves open-set LID performance.

Noisy LID is the third task, where the test set involves noisy utterances of 5 languages (Cantonese, Japanese, Russian, Korean and Mandarin). This challenge was settled for an important real-life issue of speech technology, i.e., dealing with the low SNR condition.

The remainder of this paper is organized as follows. Section 2 presents the data profile. Section 3 gives an introduction to the evaluation metrics, as well as the baselines. Section 4 analyses the results in detail and summarizes the methodologies involved in the competition. Section 5 discusses some novel approaches. In section 6, we make a conclusion on OLR 2020 Challenge.

\section{Database profile}

The OLR 2020 Challenge ﻿involved 16 languages and 3 Chinese dialects, provided by Speechocean and the NSFC M2ASR project\cite{wang2017m2asr}, and all the data is free for participants. The datasets of OLR 2020 Challenge include the data from the past four OLR challenges and two newly provided datasets. ﻿﻿All the data that could be used for submission systems ﻿construction are listed below.

\begin{itemize}
\item AP16-OL7: The standard database for AP16-OLR, including AP16-OL7-train, AP16-OL7-dev and AP16-OL7-test.
\item AP17-OL3: A dataset provided by the M2ASR project, involving three additional languages. It contains AP17-OL3-train and AP17-OL3-dev.
\item AP17-OLR-test: The standard test set for AP17-OLR. It contains AP17-OL7-test and AP17-OL3-test.
\item AP18-OLR-test: The standard test set for AP18-OLR. It contains AP18-OL7-test and AP18-OL3-test.
\item AP19-OLR-dev: The development set for AP19-OLR. It contains AP19-OLR-dev-task2 and AP19-OLR-dev-task3.
\item AP19-OLR-test: The standard test set for AP19-OLR. It contains AP19-OL7-test and AP19-OL3-test.
\item AP20-OLR-dialect: The newly provided training set, including three kinds of Chinese dialects.
\item THCHS30: The THCHS30 database (plus the accompanied resources) published by CSLT, Tsinghua University\cite{2015THCHS}.
\end{itemize}

%The test dataset AP20-OLR-test includes three parts corresponding to the three LID tasks, had been provided to Participants after the submission deadline. More informations about these datasets can be found in the ﻿“AP20-OLR Challenge: Three Tasks and Their Baselines”\cite{li2020ap20}.

The test dataset AP20-OLR-test includes three parts, corresponding to the three LID tasks, respectively. The labelled test set was provided to participants after the submission deadline. More informations about these datasets can be found in\cite{li2020ap20}.

\begin{figure}[h]
  \centering
  \includegraphics[width=\linewidth]{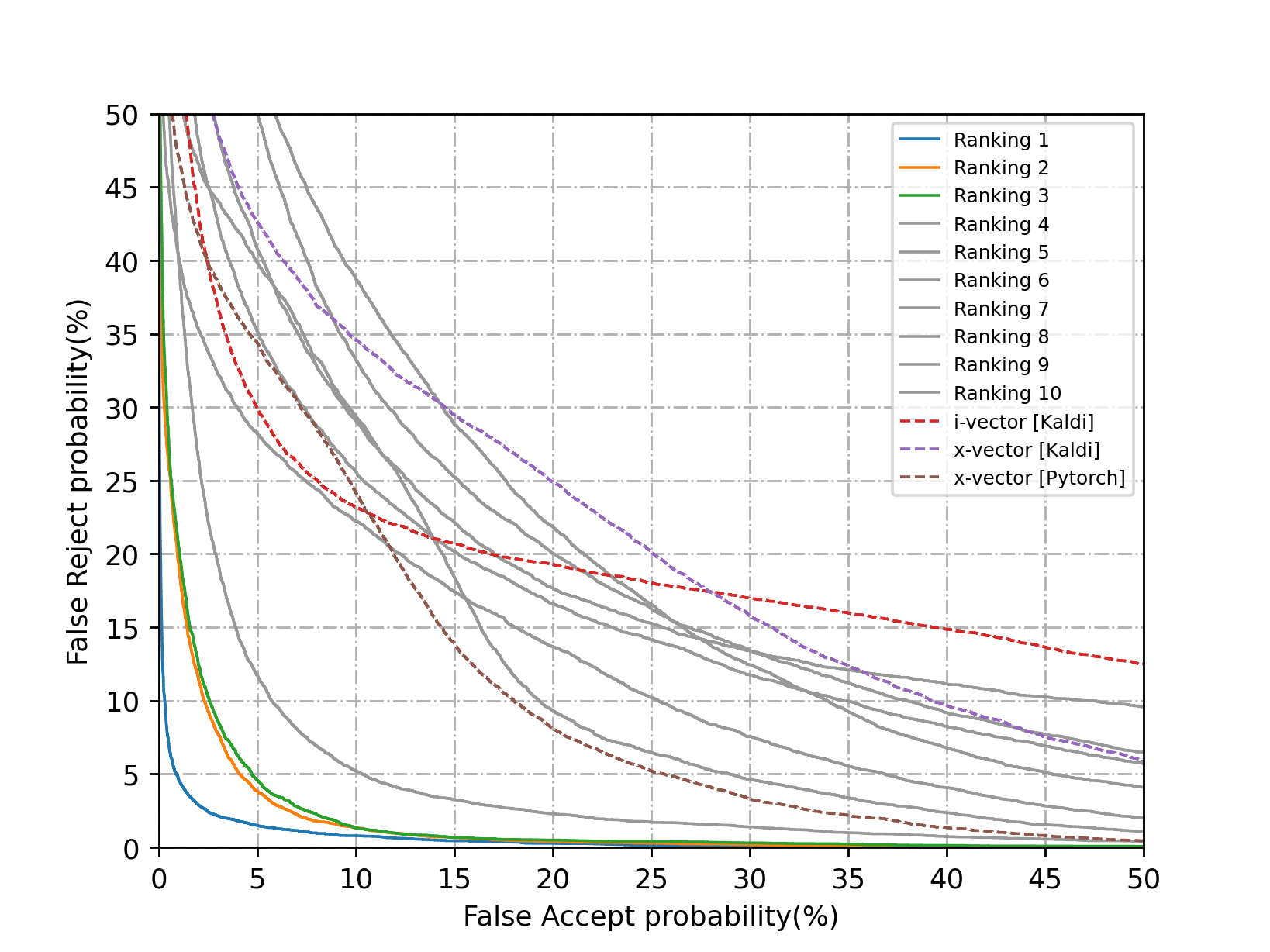}
  \caption{DET curves for Cross-channel LID.}
  \label{fig:speech_production}
\end{figure}

\begin{figure}[h]
  \centering
  \includegraphics[width=\linewidth]{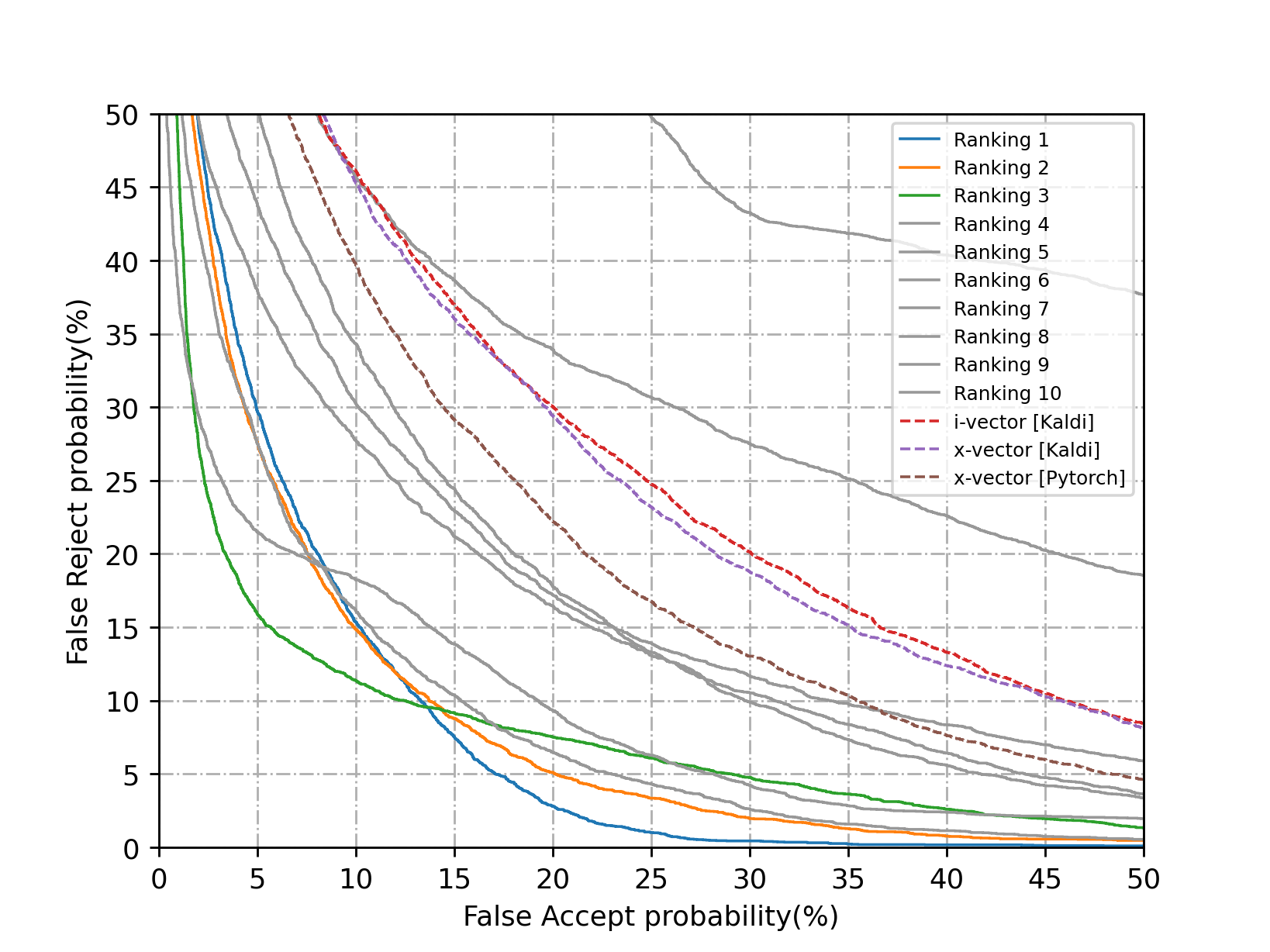}
  \caption{DET curves for Open-set dialect identification.}
  \label{fig:speech_production}
\end{figure}

\begin{figure}[th]
  \centering
  \includegraphics[width=\linewidth]{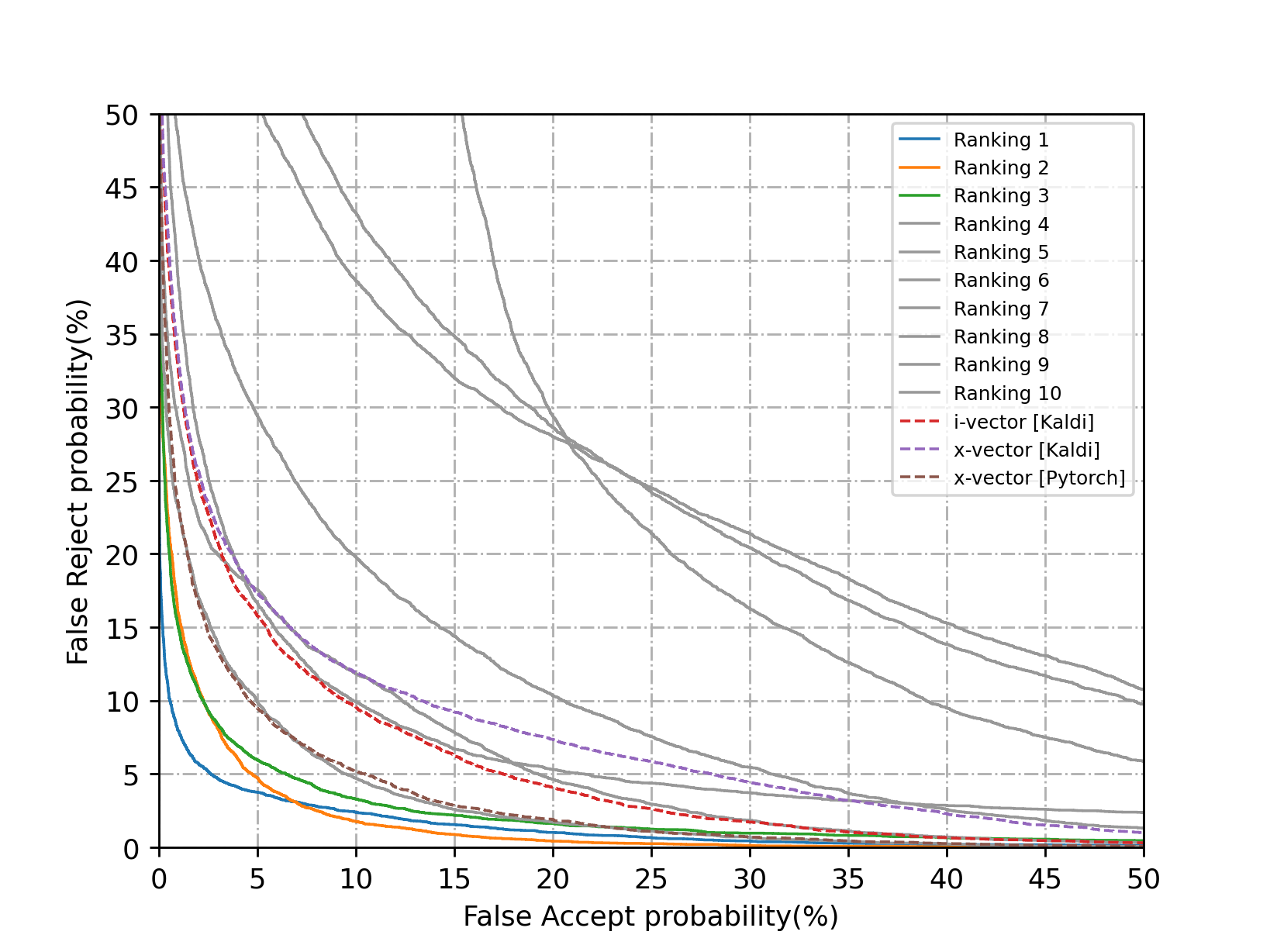}
  \caption{DET curves for Noisy LID.}
  \label{fig:speech_production}
\end{figure}

\section{Evaluation metrics and baselines}
As in NIST LRE15\cite{martin2015nist}, the OLR 2020 Challenge chooses $C_{avg}$ as the principle evaluation metric, and the Equal Error Rate (EER) as the secondary metric. First define the pair-wise loss that composes the missing and false alarm probabilities for a particular target/non-target language pair:

\begin{equation}
\begin{aligned}
C\left(L_{t}, L_{n}\right)=&P_{Target } P_{Miss } \left(L_{t}\right)+\left(1-P_{Target }\right) P_{F A}\left(L_{t}, L_{n}\right)
\end{aligned}
\end{equation}
where $L_{t}$ and $L_{n}$ are the target and non-target languages, respectively; $P_{Miss }$ and $P_{F A}$ are the missing and false alarm probabilities, respectively. $P_{Target}$ is the prior probability for the target language, which is set to 0.5 in the evaluation. Then the principle metric $C_{a v g}$ is defined as the average of the above pair-wise performance:

\begin{equation}
   C_{avg} = \frac{1}{N} \sum_{L_t} \left\{
  \begin{aligned}
    &\ P_{Target} \cdot P_{Miss}(L_t) \\
    &+\sum_{L_n}\ P_{Non-Target} \cdot P_{FA}(L_t, L_n)\
  \end{aligned}
  \right\}
\end{equation}
where $N$ is the number of languages, and $P_{Non-Target}=\left(1-P_{Target}\right)/(N-1)$. For the open-set testing condition, all of the interfering languages are treated as a single unknown language in the computation of $C_{a v g}$. We have provided the evaluation scripts for system development.

%\subsection{Baseline systems}
Two kinds of baseline LID systems were constructed in this challenge: the i-vector model baseline and the extended TDNN (E-TDNN)\cite{etdnn} x-vector model baseline, respectively. The feature extracting and back-end were all conducted with Kaldi. We built the x-vector model baselines with Kaldi\cite{2012The} and Pytorch\cite{paszke2019pytorch,subtools}, respectively. To provide more options, we also built the i-vector model baseline with Kaldi. The recipes of these baselines can be downloaded from the project web site\footnote{https://github.com/Snowdar/asv-subtools\#2-ap-olr-challenge-2020-baseline-recipe-language-identification}.

\section{Challenge results}
\subsection{System descriptions}
Figure 1, 2, and 3 illustrate the Detection Error Tradeoff (DET) curves for the top 10 systems and 3 baselines on three tasks, where the colored solid line, gray solid line, and colored dotted line represent the top 3 systems, other systems, and baseline systems on Kaldi or Pytorch, respectively. 

For the three tasks, the optimal baseline is the x-vector system built with Pytorch, and all the top 3 systems outperform the optimal baseline system with a large margin. For Task 1, the top 3 systems deliver EERs below 5\%. Most of the systems deliver EERs between 15\% and 20\%. The top 3 systems of Task 2 deliver EERs between 10\% and 15\%, with a small number of systems outside of 25\%. The EERs of the top 3 systems in Task 3 were around 5\%, and the other systems were below 25\%. It can be seen that the EERs of the top 3 systems for both Task 1 and Task 3 are already at a lower level, while for the open-set testing condition, the EERs of the top 3 systems are relatively higher, indicating that the open-set testing is still a difficult task for language recognition. 
Descriptions of the top 3 systems for each task can be downloaded from the challenge web site\footnote{http://cslt.riit.tsinghua.edu.cn/mediawiki/index.php?title=OLR\_Challenge\_2020}.
%加脚释\footnote{http://cslt.riit.tsinghua.edu.cn/mediawiki/index.php/OLR_Challenge_2020}

\subsubsection{Cross-channel LID}
$C_{avg}$ and EERs for the top 10 teams on the cross-channel LID task are shown in Table 1. For this task, 17 teams in total submitted valid scores, with 4 teams outperforming the Pytorch baseline. The champion system, LORIA-Inria-Multispeech, achieved a $C_{avg}$ of 0.0239 and an EER of 2.47\%. They used embeddings extracted from an intermediate bottleneck layer of a multilingual Deep Neural Network (DNN), trained with the Connectionist Temporal Classification (CTC) loss\cite{ctc}. These bottleneck embeddings were used as input features to train a DNN LID system. They tried different loss functions such as Additive Angular Margin softmax (AAM-softmax)\cite{aam}, Maximum Mean Discrepancy (MMD)\cite{mmd} and n-pair loss function\cite{n-pair}. They chose Stochastic Gradient Descent (SGD)\cite{sgd} to optimize the LID model, and average the parameters of candidate models by Stochastic Weight Average (SWA)\cite{swa}. Finally, they fused their subsystems by the Focal toolkit\footnote{https://sites.google.com/site/nikobrummer/focalmulticlass}. The team at the second place used semi-supervised training and the model was based on x-vector, while the third place team chose to use Residual Network (ResNet)\cite{rsnet} with a deformation structure of the Gated Recurrent Unit (GRU).

%Task1: Cross-channel LID Top 10
\begin{table}[t]
\centering
\caption{Cross-channel LID Top 10. \# indicates baselines on Kaldi and Pyorch, respectively.}
\begin{tabular}{cccc} 
\toprule
 \textbf{Ranking}  & \textbf{Team Name}      & \textbf{$C_{avg}$}              & \textbf{EER (\%)}               \\ 
\midrule
1                  & LORIA-Inria-Multispeech & 0.0239                     & 2.47                       \\
2                  & NTU-XJU                 & 0.0421                     & 4.51                       \\
3                  & Malaxiaolongxia         & 0.0477                     & 4.82                       \\
4                  & IBG\_AI                 & 0.0760                     & 7.51                       \\
\#                 & x-vector ﻿[Pytorch]     & \multicolumn{1}{l}{0.1321}   & {15.48}  \\
5                  & youdao                  & 0.1377                     & 15.99                      \\
6                  & RoyalFlush              & 0.1483                     & 16.37                      \\
\#                 & i-vector ﻿[Kaldi]       & 0.1542                     & 19.40                      \\
7                  & gz                      & 0.1669                     & 18.00                      \\
8                  & Phonexia                & 0.1694                     & 18.70                      \\
9                  & BJFU                    & 0.2037                     & 20.86                      \\
10                 & Anonymous               & 0.2088                     & 20.03                      \\
\#                 & x-vector ﻿[Kaldi]       & 0.2098                     & 22.49                      \\
\bottomrule
\end{tabular}
\end{table}

%Task2: Open-set dialect identification Top 10
\begin{table}[t]
\centering
\caption{Open-set dialect identification Top 10. \# indicates baselines on Kaldi and Pyorch, respectively.}
\begin{tabular}{cccc} 
\toprule
 \textbf{Ranking}  & \textbf{Team Name}  & \textbf{$C_{avg}$}  & \textbf{EER (\%)}   \\ 
\midrule
1                  & Phonexia            & 0.0738         & 11.97          \\
2                  & Royal-Flush         & 0.0871         & 11.97          \\
3                  & IBG\_AI             & 0.1096         & 10.84          \\
4                  & youdao              & 0.1116         & 14.42          \\
5                  & Anonymous           & 0.1312         & 12.67          \\
6                  & NTU-XJU             & 0.1546         & 18.02          \\
7                  & BJFU                & 0.1808         & 18.60          \\
\#                 & x-vector ﻿[Pytorch] & 0.1938         & 19.74          \\
8                  & gz                  & 0.2084         & 19.08          \\
\#                 & x-vector ﻿[Kaldi]   & 0.2370         & 22.25          \\
\#                 & i-vector ﻿[Kaldi]   & 0.2439         & 23.94          \\
9                  & Anonymous           & 0.2614         & 28.40          \\
10                 & ABSPlab\_IIT\_KGP   & 0.3848         & 40.32          \\
\bottomrule
\end{tabular}
\end{table}

\subsubsection{Open-set dialect identification}
Table 2 shows the $C_{avg}$ and EERs for the top 10 systems and the baselines. For this task, among the 13 participating teams, 7 teams had better results than the best baseline. The top-performing system, Phonexia, achieved a $C_{avg}$ of 0.0738 and an EER of 11.97\%. They trained ResNet18 by Pytorch with SGD optimizer. The model was trained with the Cross Entropy (CE) loss. The input to the ResNet18 was on raw features (i.e. no VAD, no mean-variance normalization). The third-place team also chose the ResNet architecture, including 5 ResNet subsystems. Among them, the optimal single system on the development set is ResNet18 plus Sequeze and Excitation (SE) block\cite{se}, which is the same as ResNet18 selected by the Top 1 system, corroborating that ResNet18 has a certain advantage in open-set dialect identification. Too deep or too shallow network structures are all suboptimal. To reduce the impact of open-set testing condition, the second place system first trained a transformer-based CTC loss or attention end-to-end Automatic Speech Recognition (ASR) model based on the ESPNet platform\cite{espnet}. Then, they trained a 6-layer transformer as the LID model to classify the three dialects. The training of transformer uses knowledge transfer learning by using the 12-layer encoder of the transformer-based CTC or attention ASR model. Because of the materialization of the phonetic information, the recognition accuracy of the open-set testing is improved.

%加入第二名

\subsubsection{Noisy LID}
The $C_{avg}$ and EERs of the top 10 teams and baselines are presented in Table 3. Four out of fourteen submitted scores outperformed the Pytorch x-vector baseline. The top 1 system, LORIA-Inria-Multispeech, achieved a $C_{avg}$ of 0.0374 and an EER of 4.07\%. Their submitted system is the same as the one for Task 1, with data augmentation to attenuate the effect of noise, e.g., adding white noise and babble noise (generated by mixing other files from the training set) and convolving with random artificial band-pass filters. The second-place team also submitted the same system as in Task 1, and on top of that, the enrollment set of Task 3 was augmented with background noise to reduce the effect of noise. They extracted noise and all data of randomly chosen 4 speakers for each target languages (Cantonese, Japanese, and Mandarin) from AP16-OL7, and merged noisy part of Korean and Russian data into them. Finally, they combined those merged noisy data with noisy part of Korean and Russian data as enrollment set of Task 3.

% •表示放置两栏
\begin{figure*}[t]
  \centering
  \includegraphics[width=\linewidth]{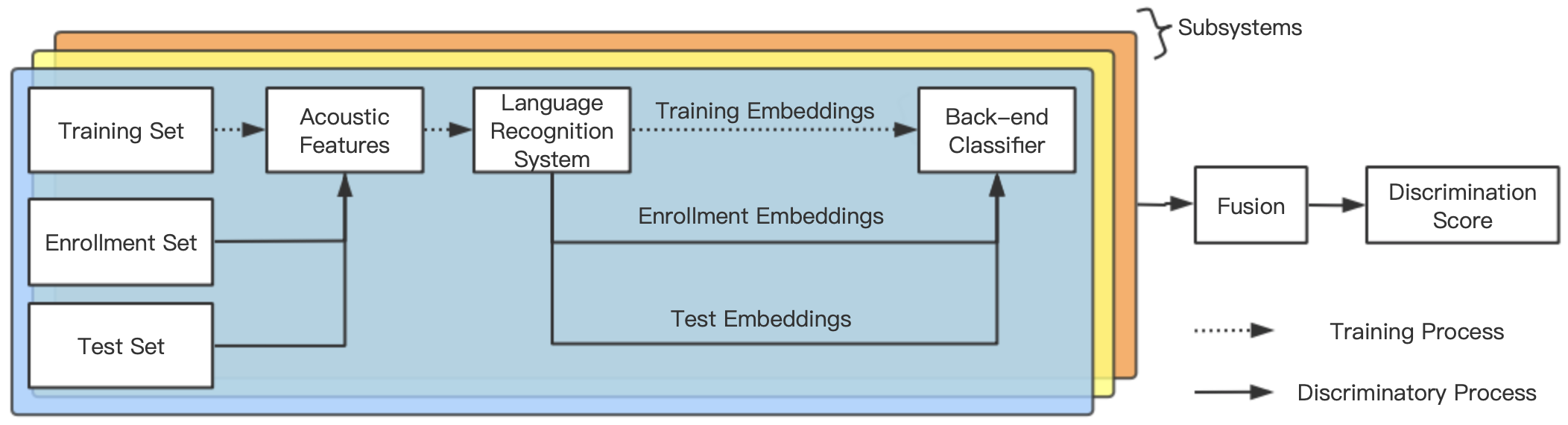}
  \caption{The flow chart of language recognition.}
  \label{fig:speech_production}
\end{figure*}

%Task3: Noisy LID Top 10
\begin{table}
\centering
\caption{Noisy LID Top 10. \# indicates baselines on Kaldi and Pyorch, respectively.}
\begin{tabular}{cccc} 
\toprule
\textbf{Ranking} & \textbf{Team Name}      & \textbf{$C_{avg}$} & \textbf{EER (\%)}  \\ 
\midrule
1                & LORIA-Inria-Multispeech & 0.0374        & 4.07          \\
2                & NTU-XJU                 & 0.0476        & 4.87          \\
3                & Malaxiaolongxia         & 0.0538        & 5.60          \\
4                & RoyalFlush              & 0.0722        & 7.12          \\
\#               & x-vector ﻿[Pytorch]     & 0.0715        & 7.14          \\
\#               & i-vector ﻿[Kaldi]       & 0.0967        & 9.77          \\
5                & Phonexia                & 0.0986        & 9.96          \\
6                & youdao                  & 0.1057        & 11.10         \\
\#               & x-vector ﻿[Kaldi]       & 0.1079        & 11.12         \\
7                & gz                      & 0.1467        & 14.70         \\
8                & BJFU                    & 0.2261        & 23.40          \\
9                & Anonymous               & 0.2326        & 24.69         \\
10               & Anonymous               & 0.2444        & 24.62         \\
\bottomrule
\end{tabular}
\end{table}

\subsection{Technology summary}
Figure 4 shows the flow of language recognition. We will summarize the acoustic features, language recognition models, back-end classifiers and score fusion strategies that OLR 2020 teams have used in their submissions, as described below.

\begin{itemize}[leftmargin= 9 pt]
\item Input features: Augmentation (e.g., velocity, volume perturbations) was widely used. Some systems used background noise extracted from the training data, white noise and random artificial band-pass filters. Most systems applied the SpecAugment strategy\cite{spec} during training. Mel-filterbank (FBank) and  Mel Frequency Cepstral Coefficients (MFCCs) were two most frequently used acoustic features. A few systems used Perceptual Linear Predictive (PLP)\cite{plp} and spectrum as acoustic features.
\item Structure and Optimization: The standard TDNN and ResNet architecture were two mostly used model structures. Some of them used other network structures, such as Convolutional Neural Networks (CNN), incorporating the SE blocks in the deep Residual Networks (ResNet-SE), Polar Transformer Network (PTN)\cite{ptn}, GRU, BLSTM\cite{blstm}, attention structure, attentive pooling, Global Context Network (GCNet)\cite{gcnet}, NetVLAD\cite{netvlad} or inspired Vector of Locally Aggregated Descriptors (VLAD)\cite{vlad}.
\item Auxiliary information: The introduction of ASR to help language recognition was investigated by top teams (two out of five top teams used E2E ASR technologies). The phonetic information based on spectral features and Bottleneck Features (BNFs) or multi-task learning can be further investigated to improve the performance of LID.
\item Loss: CE and Additive Margin softmax (AM-softmax) were two mostly used loss functions. AAM-softmax, MMD and n-pair loss were also experimented with.
\item Scoring backend: After the LDA projection and centering, the Logistic Regression (LR) was used to compute the score; one team used Gaussian Mixture Model (GMM) to enhance the performance. Cosine, Support Vector Machines (SVM) and PLDA were also chosen.
\item Model fusion: Most of submitted systems were the fusion of many subsystems. The fusion was mostly based on the score, and the fusion methods included average fusion, greedy fusion and the fusion approach implemented in the FoCal toolkit.
\item Platform: Most teams chose to use ASV-Subtools (Pytorch)\cite{subtools}, Kaldi, Pytorch; a few teams chose ESPNet, TensorFlow, Matlab or Web Real-Time Communication (WebRTC).
\end{itemize}

\section{Discussion}
In this challenge, most teams used the x-vector system with the E-TDNN structure as the baseline and used LR as the back-end. One interesting finding was that multiple teams submitted the same system on Task 1 and Task 3, and the same system received good performance in both tasks, e.g., the top 1 system on Task 1 had a $C_{avg}$ of 0.0239 and an EER of 2.47\%, and the same system obtained a $C_{avg}$ of 0.0374 and an EER of 4.07\% on Task 3. Because cross-channel and noise can be regarded as two kinds of distortion in the background environment, many systems cast them into one problem, i.e cross-domain. To counteract this distortion, some systems added background noise extracted from other datasets to the training set and enrollment set and obtained phonetic information from the speech recognition system. For the open-set testing condition, many teams chose the ResNet network structure coupled with SGD optimizer, which allows to use very deep networks with residual connections. In summary, it can be seen that the utilization of phonetic  information and the ResNet structure contribute the most to language recognition in OLR 2020 challenge. In the future, involving more auxiliary or multimodal information and investigating more suitable network structures may be the most promising directions for language recognition.

\section{Conclusions}
In this paper, we presented the database profile, evaluation metrics, task definitions and their baselines of OLR 2020 Challenge. We also summarized the ranking results of the participating teams and analyzed the technologies that the participants used, as well as language recognition. This competition was designed based on realistic scenarios with three tasks: (1) cross-channel LID, (2) open-set dialect identification, and (3) noisy LID. Due to public availability of the script for the baseline systems, more people participate in this competition. The participants used a wide variety of approaches to improve the performance on the three tasks, especially by obtaining additional information through ASR model to reduce the impact of noise, unknown channels, and unknown test sets.
\section{Acknowledgements}
This work was supported by the National Natural Science Foundation of China No.61876160, No.61633013 and No.62001405.

We would like to thank Ming Li at Duke Kunshan University, Xiaolei Zhang at Northwestern Polytechnical University for their help in organizing this OLR 2020 challenge.

\bibliographystyle{IEEEtran}

\bibliography{ap20-olr}

\begin{thebibliography}{10}
\providecommand{\url}[1]{#1}
\csname url@samestyle\endcsname
\providecommand{\newblock}{\relax}
\providecommand{\bibinfo}[2]{#2}
\providecommand{\BIBentrySTDinterwordspacing}{\spaceskip=0pt\relax}
\providecommand{\BIBentryALTinterwordstretchfactor}{4}
\providecommand{\BIBentryALTinterwordspacing}{\spaceskip=\fontdimen2\font plus
\BIBentryALTinterwordstretchfactor\fontdimen3\font minus
  \fontdimen4\font\relax}
\providecommand{\BIBforeignlanguage}[2]{{%
\expandafter\ifx\csname l@#1\endcsname\relax
\typeout{** WARNING: IEEEtran.bst: No hyphenation pattern has been}%
\typeout{** loaded for the language `#1'. Using the pattern for}%
\typeout{** the default language instead.}%
\else
\language=\csname l@#1\endcsname
\fi
#2}}
\providecommand{\BIBdecl}{\relax}
\BIBdecl

\bibitem{wang2016ap16}
D.~Wang, L.~Li, D.~Tang, and Q.~Chen, ``Ap16-ol7: A multilingual database for
  oriental languages and a language recognition baseline,'' in \emph{2016
  Asia-Pacific Signal and Information Processing Association Annual Summit and
  Conference (APSIPA)}.\hskip 1em plus 0.5em minus 0.4em\relax IEEE, 2016, pp.
  1--5.

\bibitem{tang2017ap17}
Z.~Tang, D.~Wang, Y.~Chen, and Q.~Chen, ``Ap17-olr challenge: Data, plan, and
  baseline,'' in \emph{2017 Asia-Pacific Signal and Information Processing
  Association Annual Summit and Conference (APSIPA ASC)}.\hskip 1em plus 0.5em
  minus 0.4em\relax IEEE, 2017, pp. 749--753.

\bibitem{2018AP18}
Z.~Tang, D.~Wang, and Q.~Chen, ``Ap18-olr challenge: Three tasks and their
  baselines,'' in \emph{2018 Asia-Pacific Signal and Information Processing
  Association Annual Summit and Conference (APSIPA ASC)}, 2018.

\bibitem{tang2019ap19}
Z.~Tang, D.~Wang, and L.~Song, ``Ap19-olr challenge: Three tasks and their
  baselines,'' in \emph{2019 Asia-Pacific Signal and Information Processing
  Association Annual Summit and Conference (APSIPA ASC)}.\hskip 1em plus 0.5em
  minus 0.4em\relax IEEE, 2019, pp. 1917--1921.

\bibitem{li2020ap20}
Z.~Li, M.~Zhao, Q.~Hong, L.~Li, Z.~Tang, D.~Wang, L.~Song, and C.~Yang,
  ``Ap20-olr challenge: Three tasks and their baselines,'' in \emph{2020
  Asia-Pacific Signal and Information Processing Association Annual Summit and
  Conference (APSIPA ASC)}.\hskip 1em plus 0.5em minus 0.4em\relax IEEE, 2020,
  pp. 550--555.

\bibitem{izenman2013linear}
A.~J. Izenman, ``Linear discriminant analysis,'' in \emph{Modern multivariate
  statistical techniques}.\hskip 1em plus 0.5em minus 0.4em\relax Springer,
  2013, pp. 237--280.

\bibitem{ioffe2006probabilistic}
S.~Ioffe, ``Probabilistic linear discriminant analysis,'' in \emph{European
  Conference on Computer Vision}.\hskip 1em plus 0.5em minus 0.4em\relax
  Springer, 2006, pp. 531--542.

\bibitem{wang2017m2asr}
D.~Wang, T.~F. Zheng, Z.~Tang, Y.~Shi, L.~Li, S.~Zhang, H.~Yu, G.~Li, S.~Xu,
  A.~Hamdulla \emph{et~al.}, ``M2asr: Ambitions and first year progress,'' in
  \emph{2017 20th Conference of the Oriental Chapter of the International
  Coordinating Committee on Speech Databases and Speech I/O Systems and
  Assessment (O-COCOSDA)}.\hskip 1em plus 0.5em minus 0.4em\relax IEEE, 2017,
  pp. 1--6.

\bibitem{2015THCHS}
D.~Wang and X.~Zhang, ``Thchs-30 : A free chinese speech corpus,''
  \emph{Computer Science}, 2015.

\bibitem{martin2015nist}
A.~F. Martin, C.~S. Greenberg, J.~M. Howard, D.~Bans{\'e}, G.~R. Doddington,
  J.~Hern{\'a}ndez-Cordero, and L.~P. Mason, ``Nist language recognition
  evaluation—plans for 2015,'' in \emph{Sixteenth Annual Conference of the
  International Speech Communication Association}, 2015.

\bibitem{etdnn}
D.~Snyder, D.~Garcia-Romero, G.~Sell, A.~Mccree, and D.~Povey, ``Speaker
  recognition for multi-speaker conversations using x-vectors,'' in
  \emph{ICASSP 2019 - 2019 IEEE International Conference on Acoustics, Speech
  and Signal Processing (ICASSP)}, 2019.

\bibitem{2012The}
D.~Povey, A.~Ghoshal, G.~Boulianne, L.~Burget, O.~Glembek, N.~Goel,
  M.~Hannemann, P.~Motlíček, Y.~Qian, P.~Schwarz, J.~Silovský, G.~Stemmer,
  and K.~Vesel, ``The kaldi speech recognition toolkit,'' \emph{IEEE 2011
  Workshop on Automatic Speech Recognition and Understanding}, 01 2011.

\bibitem{paszke2019pytorch}
A.~Paszke, S.~Gross, F.~Massa, A.~Lerer, J.~Bradbury, G.~Chanan, T.~Killeen,
  Z.~Lin, N.~Gimelshein, L.~Antiga, A.~Desmaison, A.~Köpf, E.~Yang, Z.~DeVito,
  M.~Raison, A.~Tejani, S.~Chilamkurthy, B.~Steiner, L.~Fang, J.~Bai, and
  S.~Chintala, ``Pytorch: An imperative style, high-performance deep learning
  library,'' 2019.

\bibitem{subtools}
F.~Tong, M.~Zhao, J.~Zhou, H.~Lu, Z.~Li, L.~Li, and Q.~Hong, ``{ASV-Subtools}:
  {Open} source toolkit for automatic speaker verification,'' in \emph{ICASSP
  2021-2021 IEEE International Conference on Acoustics, Speech and Signal
  Processing (ICASSP)}.\hskip 1em plus 0.5em minus 0.4em\relax IEEE, 2021, pp.
  6184--6188.

\bibitem{ctc}
A.~Graves, S.~Fernández, F.~Gomez, and J.~Schmidhuber, ``Connectionist
  temporal classification: Labelling unsegmented sequence data with recurrent
  neural 'networks,'' in \emph{ICML 2006 - Proceedings of the 23rd
  International Conference on Machine Learning}, vol. 2006, 01 2006, pp.
  369--376.

\bibitem{aam}
J.~Deng, J.~Guo, N.~Xue, and S.~Zafeiriou, ``Arcface: Additive angular margin
  loss for deep face recognition,'' in \emph{2019 IEEE Conference on Computer
  Vision and Pattern Recognition (CVPR)}, 06 2019, pp. 4685--4694.

\bibitem{mmd}
R.~Duroselle, D.~Jouvet, and I.~Illina, ``Unsupervised regularization of the
  embedding extractor for robust language identification,'' in \emph{Odyssey
  2020 The Speaker and Language Recognition Workshop}, 11 2020, pp. 39--46.

\bibitem{n-pair}
\BIBentryALTinterwordspacing
------, ``{Metric Learning Loss Functions to Reduce Domain Mismatch in the
  x-Vector Space for Language Recognition},'' in \emph{Proc. Interspeech 2020},
  2020, pp. 447--451. [Online]. Available:
  \url{http://dx.doi.org/10.21437/Interspeech.2020-1708}
\BIBentrySTDinterwordspacing

\bibitem{sgd}
N.~Srivastava, G.~Hinton, A.~Krizhevsky, I.~Sutskever, and R.~Salakhutdinov,
  ``Dropout: A simple way to prevent neural networks from overfitting,''
  \emph{Journal of Machine Learning Research}, vol.~15, pp. 1929--1958, 06
  2014.

\bibitem{swa}
\BIBentryALTinterwordspacing
P.~Izmailov, D.~Podoprikhin, T.~Garipov, D.~P. Vetrov, and A.~G. Wilson,
  ``Averaging weights leads to wider optima and better generalization,''
  \emph{CoRR}, vol. abs/1803.05407, 2018. [Online]. Available:
  \url{http://arxiv.org/abs/1803.05407}
\BIBentrySTDinterwordspacing

\bibitem{rsnet}
K.~He, X.~Zhang, S.~Ren, and J.~Sun, ``Deep residual learning for image
  recognition,'' in \emph{2016 IEEE Conference on Computer Vision and Pattern
  Recognition (CVPR)}, 06 2016, pp. 770--778.

\bibitem{se}
J.~Zhou, T.~Jiang, Z.~Li, L.~Li, and Q.~Hong, ``Deep speaker embedding
  extraction with channel-wise feature responses and additive supervision
  softmax loss function,'' in \emph{Interspeech 2019}, 09 2019, pp. 2883--2887.

\bibitem{espnet}
\BIBentryALTinterwordspacing
S.~Mehta, M.~Rastegari, A.~Caspi, L.~G. Shapiro, and H.~Hajishirzi, ``Espnet:
  Efficient spatial pyramid of dilated convolutions for semantic
  segmentation,'' \emph{CoRR}, vol. abs/1803.06815, 2018. [Online]. Available:
  \url{http://arxiv.org/abs/1803.06815}
\BIBentrySTDinterwordspacing

\bibitem{spec}
D.~Park, W.~Chan, Y.~Zhang, C.-C. Chiu, B.~Zoph, E.~Cubuk, and Q.~Le,
  ``Specaugment: A simple data augmentation method for automatic speech
  recognition,'' in \emph{Interspeech 2019}, 09 2019, pp. 2613--2617.

\bibitem{plp}
Hermansky and Hynek, ``Perceptual linear predictive (plp) analysis of speech.''
  \emph{Journal of the Acoustical Society of America}, vol.~87, no.~4, pp.
  1738--1752, 1990.

\bibitem{ptn}
C.~Esteves, C.~Allen-Blanchette, X.~Zhou, and K.~Daniilidis, ``Polar
  transformer networks,'' in \emph{International Conference on Learning
  Representations}, 02 2018.

\bibitem{blstm}
B.~H. Su, S.~L. Yeh, M.~Y. Ko, H.~Y. Chen, and C.~C. Lee, ``Self-assessed
  affect recognition using fusion of attentional blstm and static acoustic
  features,'' in \emph{Interspeech 2018}, 2018.

\bibitem{gcnet}
Y.~Cao, J.~Xu, S.~Lin, F.~Wei, and H.~Hu, ``Gcnet: Non-local networks meet
  squeeze-excitation networks and beyond,'' in \emph{2019 IEEE/CVF
  International Conference on Computer Vision Workshop (ICCVW)}, 10 2019, pp.
  1971--1980.

\bibitem{netvlad}
R.~Arandjelovic, P.~Gronat, A.~Torii, T.~Pajdla, and J.~Sivic, ``Netvlad: Cnn
  architecture for weakly supervised place recognition,'' in \emph{2016 IEEE
  Conference on Computer Vision and Pattern Recognition (CVPR)}, 06 2016, pp.
  5297--5307.

\bibitem{vlad}
A.~Amory, G.~Muhammad, and H.~Mathkour, ``Deep tree net-vector of locally
  aggregated descriptor (vlad) model,'' \emph{IEEE Access}, vol.~7, pp. 1--1,
  10 2019.

\end{thebibliography}

% \begin{thebibliography}{9}
% \bibitem[1]{Davis80-COP}
%   S.\ B.\ Davis and P.\ Mermelstein,
%   ``Comparison of parametric representation for monosyllabic word recognition in continuously spoken sentences,''
%   \textit{IEEE Transactions on Acoustics, Speech and Signal Processing}, vol.~28, no.~4, pp.~357--366, 1980.
% \bibitem[2]{Rabiner89-ATO}
%   L.\ R.\ Rabiner,
%   ``A tutorial on hidden Markov models and selected applications in speech recognition,''
%   \textit{Proceedings of the IEEE}, vol.~77, no.~2, pp.~257-286, 1989.
% \bibitem[3]{Hastie09-TEO}
%   T.\ Hastie, R.\ Tibshirani, and J.\ Friedman,
%   \textit{The Elements of Statistical Learning -- Data Mining, Inference, and Prediction}.
%   New York: Springer, 2009.
% \bibitem[4]{YourName17-XXX}
%   F.\ Lastname1, F.\ Lastname2, and F.\ Lastname3,
%   ``Title of your INTERSPEECH 2021 publication,''
%   in \textit{Interspeech 2021 -- 20\textsuperscript{th} Annual Conference of the International Speech Communication Association, September 15-19, Graz, Austria, Proceedings, Proceedings}, 2020, pp.~100--104.
% \end{thebibliography}

\end{document}